\documentclass[aps,floatfix,showpacs]{revtex4}
\usepackage{graphicx}
\usepackage{amsmath,amssymb}
\usepackage[english,activeacute] {babel}

\begin{document}
%\preprint{TAN-FNT-03-xx}

\title{Phenomenological aspects of the exotic $T$ quark in 331 models}
% \rule{0cm}{.6cm}}

\author{J.M.\ Cabarcas$^a$, D.\ G\'omez Dumm$^a$ and R.\ Martinez$^b$}

\affiliation{ $^a$ IFLP, CONICET -- Dpto.\ de F\'{\i}sica, Univ.\
Nac.\ de La Plata, \\ C.C. 67, 1900 La Plata, Argentina. \\
$^b$ Dpto.\ de F\'{\i}sica, Universidad Nacional, Bogot\'a, Colombia.}

\begin{abstract}
In the context of 331 models we analyze the phenomenology of exotic $T$
quarks with electric charge 2/3. We establish bounds for the corresponding
masses and mixing angles and study the decay modes $T\to bW$, $tZ$ and
$qH$. It is found that the decays into scalars are strongly dependent on
the model parameters, and can be the dominant ones in a scenario with
approximate flavor symmetry.
\end{abstract}


\maketitle

\section{Introduction}

In addition to the main goal of identifying a light Higgs
boson~\cite{lhiggs}, an important challenge of the LHC is the observation
of clear signals of physics beyond the Standard Model (SM). In fact, the
general consensus is that the SM is not the ultimate theory for strong and
electroweak interactions, and many models have been proposed throughout
the last three decades attempting to solve existing theoretical puzzles
such as hierarchy problems, replication of fermion families, coupling
unification, etc~\cite{models}. Since most of these models include new
physics already at the TeV scale, it is likely that the corresponding
effects could be observed at the LHC at 100~fb$^{-1}$ luminosity. In
general, the various theories of new physics predict the presence of new
(``exotic'') fermions, gauge bosons and scalar bosons.

In particular, many models include exotic $T$ quarks with electromagnetic
charge $2/3$~\cite{top}. This is e.g.\ the case of little Higgs
theories~\cite{little}, in which an extra $T$ is introduced in order to
cancel the quadratic divergence in the Higgs selfenergy coming from the
ordinary top quark, and the case of theories including extra dimensions,
in which one has towers of quark singlets $T^{(n)}_{L,R}$ that can be
associated to the left and right handed components of the SM top
quark~\cite{extra}. Extra $T$ quarks are also predicted in the context of
``331'' models, in which the standard SU(2)$_L\, \otimes\, {\rm U(1)}_Y$
electroweak gauge symmetry is enlarged to SU(3)$_L\, \otimes\, {\rm
U(1)}_X$~\cite{pleitez}. In these models, extra fermions have to be added
to the ordinary SM quarks in order to complete the corresponding SU(3)$_L$
triplets. In the LHC, a pair of these exotic fermions can be produced
through gluon fusion or quark-antiquark annihilation~\cite{TT} in the
reaction $pp\to T\bar{T}$, or a single $T$ quark can come out through the
reaction $bq\to Tq'$~\cite{Tj}. Estimations in
Refs.~\cite{top,AguilarSaavedra:2005pv} show that these reactions can be
distinguished in the LHC at $100~fb^{-1}$ luminosity, even for exotic
fermion masses of the order of 1 TeV. The corresponding background is
basically given by SM top quark production, offering the possibility of
finding a clear signature of new physics~\cite{Agashe:2006hk}. Thus, the
analysis of $T$ production and decay in definite models is a subject that
deserves detailed study.

In this work we concentrate on the study of $T$ phenomenology in
331-symmetric models. These schemes offer an explanation to the puzzle of
family replication, since the requirements of anomaly cancellation imply a
relation between the number of fermion families and the number of
colors~\cite{pleitez,anomalia}. In addition, in this context it is
possible to fit the observed neutrino masses and mixing
angles~\cite{Dias:2005yh}. Owing to the enlarged gauge symmetry, the
models include new gauge bosons and exotic quarks that behave as singlets
under standard SU(2)$_L$ transformations. Different classes of 331 models
include either new quarks with ordinary 2/3 and -1/3 electric charges or
even more exotic fermions with charges 5/3 and -4/3 that come together
with doubly charged gauge bosons and scalar fields (in fact, this is the
case of the original versions of the model, see Ref.~\cite{pleitez}). In
order to break the gauge symmetry it is necessary to introduce a minimal
scalar sector of two triplets (this is called the ``economical''
model~\cite{Ponce:2002sg,Dong:2006mg}), while other 331 models consider
three scalar triplets and even an additional scalar SU(3)$_L$
sextet~\cite{Tonasse:1996cx,Long:1997su,Diaz:2003dk}.

Since we are interested in studying the presence of an exotic top-like
quark, we consider here a 331 model in which fermions have ordinary
electric charges~\cite{Foot:1994ym,Long:1997su}. We take into account the
general situation of a scalar sector that includes three triplets,
analyzing the couplings of the $T$ with the gauge bosons and the
experimental constraints on the mixing angles between the $T$ and the
ordinary quarks. These bounds allow us to constrain the expected widths
for $T$ decays into $tZ$ and $bW$ states for definite values of the $T$
mass. Finally, we consider the Yukawa sector of the model, studying the
decays of the $T$ into an ordinary quark and a neutral or charged scalar
boson. The predictions for the relative magnitude of these decays depends
on several unknown model parameters. However, it is possible to get
estimations of the expected rates in the context of specific schemes for
the fermion mass matrices.

The article is organized as follows. In Sect.\ II we briefly describe the
model structure, while in Sect.\ III we study in more detail the couplings
of the exotic $T$ quark with the ordinary gauge bosons. Sect.\ IV and V
are devoted to the analysis of $T$ decays into gauge boson and scalar
channels respectively. Finally, in Sect.\ VI we sketch the main outlines
of this work.

\section{Model}

As stated, in 331 models the SM gauge group is enlarged to $SU(3)_C\otimes
SU(3)_L\otimes U(1)_X$. The fermions are organized into $SU(3)_L$
multiplets, which include the standard quarks and leptons, as well as
exotic particles. Though the criterion of anomaly cancellation leads to
some constraints in the fermion quantum numbers, still an infinite number
of 331 models is allowed~\cite{anomalia}. In general, the electric charge
can be written as a linear combination of the diagonal generators of the
group,
\begin{equation}
Q\ = \ T_3\,+\,\beta\, T_8\,+\,X\ ,
\end{equation}
where $\beta$ is a parameter that characterizes the specific 331 model
particle structure and quantum numbers.

If one requires that the new quarks have ordinary electric charges $2/3$
and $-1/3$, the values of $\beta$ are restricted only to $\pm
1/\sqrt{3}$~\cite{anomalia,Foot:1994ym,Ochoa:2005ch}. Here we focus in the
model with $\beta =- 1/\sqrt{3}$, which includes only one extra $T$, thus
the quark mixing in the $Q=2/3$ sector is enlarged minimally. The fermion
sector is completed with two extra quarks $B_{1,2}$ with charge $-1/3$ and
a heavy neutrino associated to each lepton family. The situation is
sketched in Table~\ref{tab:espectro}, where $Q$ and $X$ quantum numbers
are explicitly indicated. Indices $i$ and $j$ run from 1 to 3, while
$m=1,2$. Thus, the standard left-handed quarks $U_{iL}$ and $D_{iL}$ are
organized in such a way that quarks belonging to the first two quark
families lie in the $\mathbf{3^\ast}$ representation of $SU(3)_L$, while
the third family lies in the $\mathbf{3}$. This leads to the presence of
tree level flavor changing neutral currents (FCNC)~\cite{dgd}. In the
lepton sector, all left-handed particles transform as triplets in the
$\mathbf{3}$ representation.

\begin{table}[tbp]
\[
\begin{tabular}{c|c|c}
%\hline\hline
 & $Q$ & $X$ \\ \hline
 %\hline
 & & \\
$q_{mL}=\left(
\begin{array}{r}
D_m \\
-U_m \\
B_m
\end{array}
\right) _L:\ \mathbf{3^*}$
 &
$\left(
\begin{array}{r}
-\frac 13 \\
\frac 23 \\
-\frac 13
\end{array}
\right) $
&
0
\\ & & \\
\hline
& & \\
$q_{3L}=\left(
\begin{array}{r}
U_3 \\
D_3 \\
T
\end{array}
\right) _L:\ \mathbf{3}$
&
$\left(
\begin{array}{r}
\frac 23 \\
-\frac 13 \\
\frac 23
\end{array}
\right) $
&
$\frac 13$
\\ & & \\
\hline
& & \\
$U_{iR}$, $D_{iR}$, $B_{mR}$, $T_R\ :\ \mathbf{1}$
&
$\frac 23$, $-\frac 13$, $-\frac 13$, $\frac 23$
&
$\frac 23$, $-\frac 13$, $-\frac 13$, $\frac 23$
\\ & & \\
\hline
& & \\
$\ell _{jL}=\left(
\begin{array}{c}
\nu _j \\
e_j \\
N_j
\end{array}
\right) _L :\ \mathbf{3}$
&
$\left(
\begin{array}{r}
0 \\
-1 \\
0
\end{array}
\right) $
&
$-\frac 13$
\\ & & \\
\hline
& & \\
$e_{jR}$, $N_{jR}\ :\ \mathbf{1}$
&
$-1$, \ 0
&
$-1$, \ 0
\\ & & \\
\hline
\end{tabular}
\]
\caption{Fermion content and $Q$ and $X$ quantum numbers for the 331 model
with $\beta =-1/\sqrt{3}$. Index $m$ runs from 1 to 2, while $i,j$ run
from 1 to 3.}
\label{tab:espectro}
\end{table}

The gauge bosons associated with the group $SU(3)_L$ lie as usual in the
adjoint representation of the group. Electric charge states can be defined
according to
\begin{equation}
W_\mu ^\alpha G_\alpha =\frac 12\left[
\begin{array}{ccc}
W_\mu ^3\,+\,\frac 1{\sqrt{3}}\,W_\mu ^8 & \sqrt{2}\,W_\mu ^{+} & \sqrt{2}\,Y_\mu ^0
\\
\sqrt{2}\,W_\mu ^{-} & -W_\mu ^3+\frac 1{\sqrt{3}}\,W_\mu ^8 & \sqrt{2}\,Y_\mu ^{-}
\\
\sqrt{2}\,\overline{Y}_\mu ^0 & \sqrt{2}\,Y_\mu ^{+} & -\frac 2{\sqrt{3}}\,W_\mu ^8
\end{array}
\right] ,
\label{3}
\end{equation}
while a neutral vector boson $B_\mu$ is associated with the $U(1)_X$
group. The fields $W_3$, $W_8$ and $B$ can be conveniently rotated into
states $A$, $Z$ and $Z'$ according to
\begin{eqnarray}
A_\mu  & = & S_W\, W_\mu ^3 \, + \, C_W \left( -\frac 1{\sqrt{3}}\,T_WW_\mu ^8\,
+\, \sqrt{1-\frac13\, T_W^2}\; B_\mu \right) ,  \nonumber \\
Z_\mu  & = & C_W\, W_\mu ^3 \, - \, S_W \left( -\frac 1{\sqrt{3}}\,T_WW_\mu ^8\,
+\, \sqrt{1-\frac 13\, T_W^2}\; B_\mu \right) ,  \nonumber \\
Z_\mu^{\prime } & = & -\sqrt{1-\frac 13\, T_W^2}\, W_\mu ^8\, -\,
\frac 1{\sqrt{3}}\, T_W\, B_\mu \ ,
\label{rotacion}
\end{eqnarray}
where $A$ and $Z$ are identified with the usual photon and $Z$ boson of
the SM. In Eq.~(\ref{rotacion}) we have introduced the Weinberg angle
$\theta_W$, defined by
\begin{equation}
S_W \ = \ \frac{\sqrt{3}g'}{\sqrt{3g^2+4g'}} \ ,
\end{equation}
where $g$ and $g'$ correspond to the coupling constants of the $SU(3)_L$
and $U(1)_X$ groups, respectively. $S_W$ stands for $\sin\theta_W$, etc.

This scheme clearly requires an enlarged scalar sector. We consider the
model that includes three scalar triplets
\begin{equation}
\chi =\left(
\begin{array}{c}
\chi _1^0 \\
\chi _2^- \\
\xi_\chi + i\zeta _\chi
\end{array}
\right)_{(X=-1/3)}\;\;\;\;
\rho =\left(
\begin{array}{c}
\rho _1^+ \\
\xi _\rho + i\zeta _\rho \\
\rho _3^+
\end{array}
\right)_{(X=2/3)}\;\;\;\;
\eta =\left(
\begin{array}{c}
\xi _\eta + i\zeta _\eta \\
\eta _2^- \\
\eta _3^0
\end{array}
\right)_{(X=-1/3)} \ ,
\label{higgs}
\end{equation}
where the $X$ quantum numbers are indicated explicitly (notice that
$\chi_1^0$ and $\eta_3^0$ are complex fields, while the remaining neutral
fields are real). The corresponding scalar potential is given in
Ref.~\cite{Diaz:2003dk}. The breakdown of the electroweak symmetry
proceeds in two steps: firstly, the $SU(3)_L \otimes U(1)_X$ group is
broken to $SU(2)_L\otimes U(1)_Y$ through a nonzero vacuum expectation
value (VEV) $\nu_\chi$ of the field $\xi_\chi$. This induces the (heavy)
masses of the exotic fermions and gauge bosons. Secondly, the VEVs
$\nu_\rho$, $\nu_\eta$ of the neutral fields $\xi_\rho$ and $\xi_\eta$ in
the $\rho$ and $\eta$ triplets break the symmetry to $U(1)_{\rm em}$,
providing masses to the standard quarks, leptons and gauge bosons. We
assume that there is a hierarchy between the first and second breakdown
scales, i.e.\
\begin{equation}
\nu _\chi \ \gg \ \nu _\rho\, , \ \nu _\eta \ .
\label{hierarchy1}
\end{equation}
As stated, $\nu_\rho$ and $\nu_\eta$ are responsible for the $W^\pm$ and
$Z$ boson masses, thus $\nu_\rho^2 + \nu_\eta^2 \, \simeq \,
\sqrt{2}/4\,G_F \, \simeq \, (175 \ {\rm MeV})^2$.

The approximate scalar mass eigenstates and their corresponding masses are
sketched in Table~\ref{tab2}. It can be seen that the states $\xi_\chi^0$,
$\eta _3^0$ and $\rho_3^+$ are heavy scalars, with masses of the order of
the large scale $\nu_\chi$, while $h^0$ is a light scalar that can be
associated with the SM Higgs boson. The remaining physical scalars $H^0$,
$A^0$ and $H^\pm$ have masses of the order of $\sqrt{f\nu_\chi}$, where
$f$ is a dimensionful parameter that drives a trilinear coupling
$\epsilon_{ijk}\chi_i\rho_j\eta_k$ in the scalar potential (we have
assumed that $|f|\leq \nu_\chi$, so as to avoid the introduction of a new
dimension scale). The mixing angle $\beta$ in Table~\ref{tab2} is given by
$\tan\theta_\beta=\nu_\rho/\nu_\eta$.
\begin{table}[tbp]
\begin{center}
\begin{tabular}{c|c|c}
%\hline
Mass eigenstate & Mass squared & Feature \\
\hline
%\hline
$G_{Z_\mu^{\prime}}\simeq -\zeta _\chi $ & $0$ &
$Z_\mu ^{\prime }$ Goldstone
\\ \hline
$G_{Z_\mu}\simeq S_\beta \zeta _\rho -C_\beta \zeta _\eta $ & $0$ &
$Z_\mu $ Goldstone
\\ \hline
$G_{W_\mu^\pm}^{\pm }=S_\beta \rho _1^{\pm }-C_\beta \eta _2^{\pm }$ & $0$
&
$W_\mu ^{\pm }$ Goldstone
\\ \hline
$G_{K_\mu ^{0}}^{0}\simeq -\chi _1^{0}$ & $0$ &
$K_\mu ^{0}$ Goldstone
\\ \hline
$G_{K_\mu^{\pm}}^{\pm}\simeq -\chi _2^{\pm}$ & $0$ &
$K_\mu^{\pm}$ Goldstone
\\ \hline
$h^0 \simeq S_\beta \xi _\rho +C_\beta \xi _\eta $ &
$\sim \nu_\eta^2\,$, $\nu_\rho ^2$ & SM-like scalar
\\ \hline
$A^0 \simeq C_\beta \zeta _\rho +S_\beta \zeta _\eta $ &
$\sim |f|\nu_\chi$ & physical
\\ \hline
$H^0 \simeq -C_\beta \xi _\rho +S_\beta \xi _\eta $ & $\sim
|f|\nu_\chi$ & physical
\\ \hline
$H^\pm = C_\beta \rho _1^{\pm }+S_\beta \eta _2^{\pm
}$ & $\sim |f|\nu_\chi$ & physical
\\ \hline
$\xi_\chi^0$ & $\sim \nu_\chi^2$ & physical
\\ \hline
$\eta_3^0$, $\bar\eta_3^0$ & $\sim \nu_\chi ^2$ & physical
\\ \hline
$\rho_3^\pm$ & $\sim \nu_\chi^2$ & physical
\\ \hline
\end{tabular}
\end{center}
\caption{Approximate mass eigenstates in the scalar sector.}
\label{tab2}
\end{table}

The scalars couple to fermions through Yukawa like interaction terms. In
general, these can be written as
\begin{equation}
{\cal L}_Y \ = \ \sum_{q',\Phi} \left( \sum_{m=1}^{2} h_{q'}^{m\Phi}
\, \bar q_{mL}\, q'_R \Phi \ + \ h_{q'}^{3\Phi}
\bar q_{3L}\, q'_R \Phi \right) \ + \ {\rm h.c.} \ ,
\label{yukawa-q}
\end{equation}
where the sum extends over all quarks $q'$ and scalar triplets $\Phi =
\eta,\,\rho,\,\chi$. In view of the quantum numbers in
Table~\ref{tab:espectro}, the $U(1)_X$ invariance constrain these
couplings to those that satisfy $X_{q'_R} + X_\Phi = 0$ and $X_{q'_R} +
X_\Phi = 1/3$ for the first and second term in the parentheses,
respectively. Thus the allowed combinations are $q'_R \Phi\, =\,
D_{iR}\chi^\ast, \,B_{mR}\chi^\ast,\, D_{iR}\eta^\ast,\,
B_{mR}\eta^\ast,\, U_{iR}\rho^\ast,\, T_R\rho^\ast$ for the first term,
and $q_R\Phi \, = \, D_{iR}\rho,\, B_{mR}\rho,\, U_{iR}\eta,\, T_R\eta,\,
U_{iR}\chi, T_R\chi$ for the second one. In the standard quark sector, it
can be seen that the scenario is similar to that obtained in the Two-Higgs
Doublet Model (THDM) type III~\cite{Atwood:1996vj,Cheng:1987rs}. As
expected, the nonzero VEVs of the scalar fields lead to a $4\times 4$ and
a $5\times 5$ mass matrices in the up and down quark sectors,
respectively.

\section{$T$ quark couplings to $W$ and $Z$ bosons}

As stated, in the 331 model with $\beta = - 1/\sqrt{3}$, one has two
exotic quarks $B_1$ and $B_2$ with electric charge $Q = -1/3$ and one
exotic quark $T$ with $Q = 2/3$. These nonstandard fermions can be
organized together with the ordinary up- and down-like quarks $U_i =
u,c,t$ and $D_i = b,s,d$ into fermion vectors
\begin{equation}
{\cal U}^0 \ = \ \left(
\begin{array}{c}
\\ U^0 \\ \\ T^0
\end{array}
\right) \ = \ \left(
\begin{array}{c}
u^0 \\ c^0 \\ t^0 \\ T^0
\end{array}
\right) \ , \qquad
{\cal D}^0 \ = \ \left(
\begin{array}{c}
\\ D^0 \\ \\ B^0 \!\!\!\! \begin{array}{c} \rule{0cm}{0.4cm}
\\ \rule{0cm}{0.4cm} \end{array}
\end{array}
\right) \ = \ \left(
\begin{array}{c}
d^0 \\ s^0 \\ b^0 \\ B_1^0 \\ B_2^0
\end{array}
\right) \ ,
\end{equation}
where the superindex 0 indicates that we are working with weak current
eigenstates. Using this notation, the usual SM charged weak interactions
can be written as
\begin{equation}
{\cal L}^{(cc)} \ = \ - \; \frac{g}{\sqrt{2}} \; \bar {\cal U}_{\,L}^0 \;
\gamma^\mu \; {\cal P} \; {\cal D}_L^0 \; W_\mu^+ \ + \ {\rm h.c.} \ \ ,
\label{cc}
\end{equation}
where, in order to project over the ordinary quark sector, we have
introduced a $4\times 5$ matrix ${\cal P}$ defined by
\begin{equation}
{\cal P} \ = \ \left(
\begin{array}{ccccc}
1 & 0 & 0 & 0 & 0  \\
0 & 1 & 0 & 0 & 0  \\
0 & 0 & 1 & 0 & 0  \\
0 & 0 & 0 & 0 & 0
\end{array} \right) \ .
\end{equation}
Notice that exotic quarks transform as singlets under $SU(2)_L$ transformations,
thus they do not couple with the $W^\pm$ gauge bosons.

We change now to the mass eigenstate basis ${\cal U}$, ${\cal D}$ by
introducing unitary $4 \times 4$ and $5\times 5$ rotation matrices for the
up- and down-like quark sectors, respectively:
\begin{eqnarray}
{\cal U}_L^0 = V_L^u \ {\cal U}_{L} \ \  , \  \ {\cal D}_L^0 = V_L^d \
{\cal D}_L \ .
\end{eqnarray}
It is useful to group the elements of $V_L^{u,d}$ into conveniently
defined non-quadratic submatrices,
\begin{equation}
V_L^u \ = \
\left( \begin{array}{cc}
{V_0^u}_{(3\times 3)} & {V_X^u}_{(3\times 1)} \\
{V_Y^u}_{(1\times 3)} & V_T
\end{array} \right) \qquad \ {\rm and} \qquad \
V_L^d \ = \
\left( \begin{array}{cc}
{V_0^d}_{(3\times 3)} & {V_X^d}_{(3\times 2)} \\
{V_Y^d}_{(2\times 3)} & {V_B}_{(2\times 2)}
\end{array} \right) \ .
\end{equation}
Thus, in the basis of quark mass eigenstates the couplings in Eq.~(\ref{cc}) read
\begin{eqnarray}
{\cal L}^{(cc)} & = & - \frac{g}{\sqrt{2}} \; \bar{\cal U}_L \; \gamma^\mu
\; V_L^{u\dagger} \,{\cal P}\, V_L^d \; {\cal D}_L \ W_\mu^+ \;
+ \; {\rm h.c.}
\nonumber \\
& = & -\; \frac{g}{\sqrt{2}} \; \bigg[\, \bar U_L \, \gamma^\mu V_{CKM} \,
D_L \; + \; \bar U_L \, \gamma^\mu V_0^{u\dagger} V_X^d \, B_L \; + \;
\bar T_L \, \gamma^\mu V_X^{u\dagger} V_0^d \, D_L \; + \; \bar T_L \,
\gamma^\mu V_X^{u\dagger} V^d_X \, B_L \bigg] W_\mu^+ \; + \; {\rm h.c.}\
\end{eqnarray}
Notice that the mixing matrix $V_{CKM} = V_0^{u\dagger}V_0^d$ that acts
over the SM quark sector is not in general unitary.

Owing to the enlargement of the gauge symmetry group, 331 models include
also exotic gauge bosons. In the models with $\beta = \pm 1/\sqrt{3}$
these have electric charge 0 or 1. In our case one has a heavy charged
gauge boson $Y^+$, that couples with the fermions according to
\begin{equation}
{\cal L}^{Y^+}\ = \ -\;\frac{g}{\sqrt{2}}\left(\bar u_L^0\gamma^\mu B_{1L}^0
\ + \ \bar c_L^0\gamma^\mu B_{2L}^0 \ + \ \bar T_L^0\gamma^\mu b_L^0
\right)Y_\mu^+ \ + \ {\rm h.c.} \
\end{equation}
Notice that quarks $u^0$, $c^0$ couple to $Y^+$, whereas $t^0$ does not.
This is a consequence of the structure of the model, where one of the
quark families belongs to a different $SU(3)_L$ representation than the
other two. It is also important to point out that in general $W$ and $Y$
are not mass eigenstates. They become mixed by a small mixing angle
$\theta$, which turns out to be suppressed by the weak symmetry breaking
scale ratio~\cite{Dong:2006mg}, $\theta \sim {\cal O} (\nu_\rho / \nu_\chi)$.

In particular we are interested here in the couplings that involve the
exotic quark $T$. After a redefinition of the mass states $W^+$ and $Y^+$
we obtain
\begin{eqnarray}
{\cal L}^{(cc;T)} & = & -\; \frac{g}{\sqrt{2}} \ \bar T_L \;\gamma^\mu \;
\bigg[\cos\theta\, \bigg( V_X^{u\dagger} V_0^d \, D_L \; + \;
V_X^{u\dagger} V^d_X \, B_L \bigg) \; + \; \sin\theta\, \sum_{i=1}^5 K_i \,
{\cal D}_{Li} \bigg] \; W_\mu^+ \; + \; {\rm h.c.} \ ,
\label{cc2}
\end{eqnarray}
where $K_i = V_T^\ast (V^d_L)_{3i} + V_{X1}^{u\ast} (V^d_L)_{4i} +
V_{X2}^{u\ast} (V^d_L)_{3i}$. Phenomenologically, it is well known that
$V_{CKM}$ is compatible with a unitary matrix. Therefore, it is natural to
expect the matrices $V_0^u$ and $V_0^d$ to be approximately unitary, which
implies $|V^u_{Xi}| \ll |V_T| \simeq 1$. If we also approximate
$\cos\theta$ to 1, the couplings between the $T$ quark and the ordinary
down quarks can be written as
\begin{equation}
{\cal L}^{(cc;T)} \ = \ -\; \frac{g}{\sqrt{2}} \ \bar T_L\,\gamma^\mu\,
V^{(T)}_i \, D_{Li}\; W_\mu^+ \; + \; {\rm h.c.}
\end{equation}
with the definition
\begin{equation}
V^{(T)}_i \ = \ \sum_{j=1}^3 V^{u\ast}_{Xj}\; {V_0^d}_{ji} \; + \;
V_T^\ast\,\sin\theta\; {V_0^d}_{3i}\ .
\label{vt}
\end{equation}

Let us now perform a similar analysis for the neutral currents. The
couplings between the quarks and the $Z$ boson in the 331 model with
$\beta = -1/\sqrt{3}$ read~\cite{Ochoa:2005ch,Dong:2006mg}
\begin{eqnarray}
{\cal L}^{(nc)} & = & - \,\frac{g}{2C_W} \bigg[ (1 - \frac{4}{3}
S_W^2)\,\bar U_L^0 \gamma^\mu U_L^0 \; - \; \frac{4}{3} S_W^2\, \bar T_L^0
\gamma^\mu T_L^0
\; + \; (-1 + \frac{2}{3} S_W^2)\, \bar D_L^0 \gamma^\mu D_L^0\nonumber \\
& &  \; + \; \frac{2}{3} S_W^2\, \bar B_L^0 \gamma^\mu B_L^0 \; - \;
\frac{4}{3} S_W^2\, \bar {\cal U}_R^0 \gamma^\mu {\cal U}_R^0 \; + \;
\frac{2}{3}\, S_W^2 \bar {\cal D}_R^0 \gamma^\mu {\cal D}_R^0 \bigg] Z_\mu
\nonumber \\
& = & - \,\frac{g}{2C_W} \bigg[ \bar U_L^0 \gamma^\mu U_L^0 \; - \; \bar
D_L^0 \gamma^\mu D_L^0 \; - \; \frac{4}{3} S_W^2\, \bar {\cal U}^0
\gamma^\mu {\cal U}^0 \; + \; \frac{2}{3}\, S_W^2 \bar {\cal D}^0
\gamma^\mu {\cal D}^0 \bigg] Z_\mu \ .
\end{eqnarray}
In the last expression, only the first two terms in the r.h.s.\ lead to a
mixing between current eigenstates when one rotates to the mass basis. The
interactions involving the exotic quark $T$ read then
\begin{equation}
{\cal L}^{(nc;T)} \ = \ -\; \frac{g}{2C_W} \ \bar T_L\,\gamma^\mu\,
V_X^{u\dagger} V_0^u \, U_L \; Z_\mu \; + \; {\rm h.c.}
\label{nct}
\end{equation}
As in the case of the charged states $W^+$ and $Y^+$, the $Z$ state
becomes mixed with other neutral gauge bosons. However, the interactions
between the $T$ quark and the standard up-like quarks mediated by the
exotic neutral gauge bosons are expected to suffer a twofold suppression:
on one hand, the mixing angles between the gauge bosons suffer a strong
suppression ${\cal O} (\nu_\rho^2 /
\nu_\chi^2)$~\cite{Ochoa:2005ch,Dong:2006mg}, and on the other hand, for
the neutral currents one expects a mechanism of suppression of flavor
changing currents in order to be compatible with experimental constraints
(which mainly come from the down-like quark sector). In this way, the
contributions to $T$ decay arising from this mixing will be neglected in
our analysis.

\section{Bounds for $T\to U_iZ$ and $T\to D_iW$ decay widths}

The mass of the $T$ quark is expected to be of the order of the $\nu_\chi$
scale, i.e.\ at the TeV range, therefore we are not able to establish
bounds for the mixing angles from direct $T$ production before having at
our disposal the results of forthcoming experimental devices as the LHC or
ILC~\cite{Weiglein:2004hn}. However, it is possible to set bounds for the
$T$ decay widths by taking into account contributions from virtual $T$
quarks to lower energy processes. Here we concentrate in the observables
$\Delta m_K$, $\Delta m_{B_d}$ and $\Delta m_{B_s}$, which typically lead
to the most stringent constraints for the presence of new physics. Charged
currents involving $T$ quarks will contribute to these magnitudes through
one-loop box diagrams including one or two virtual $T$'s. Moreover, these
contributions are expected to be enhanced by the large $T$ mass, just as
happens in the case of the top quark.

The theoretical expressions for the contributions to the mentioned
observables driven by box diagrams can be written as
\begin{eqnarray}
\Delta m_K & = & m_{K_L} - m_{K_S} \ = \
\frac{G_F^2}{12\pi^2}\, m_W^2\, m_K\, f_K^2\, B_K \, C_K \nonumber \\
\Delta m_{B_q} & = & m_{B_{qH}^0} - m_{B_{qL}^0} \ = \
\frac{G_F^2}{12\pi^2}\, m_W^2\, m_{B_q}\, f_{B_q}^2 B_{B_q} \left|
C_{B_q}^{(SM)} + C_{B_q}^{(T)} \right| \ , \ \ \ \ q \, =\, d \,, \, s
\label{deltam}
\end{eqnarray}
where $f_P$ are the $P$ meson weak decay constants, and the parameters
$B_P$ account for the theoretical uncertainties related with the
evaluation of matrix elements that involve hadronic
states~\cite{Gamiz:2006sq,Aoki:2003xb,Dalgic:2006gp}. The coefficients
$C_K$ and $C_{B_q}$ are given by the sum of the contributions of boxes
that include ordinary and exotic quarks. Explicitly one has
\begin{eqnarray}
C_K & = & \sum_{f,f'=u,c,t} \lambda_{f,sd}\, \lambda_{f',sd}\,
E_{\,{}_\Box}(x_f,x_{f'}) \ + \ 2 \sum_{f=u,c,t} \lambda_{f,sd} \,
\lambda'_{sd} \, E_{\,{}_\Box}(x_f,x_T) \ + \ {\lambda'_{sd}}^2
\, E_{\,{}_\Box}(x_T,x_T) \nonumber \\
C_{B_q} & = & \sum_{f,f'=u,c,t} \lambda_{f,bq}\, \lambda_{f',bq}\,
E_{\,{}_\Box}(x_f,x_{f'}) \ + \ 2 \sum_{f=u,c,t} \lambda_{f,bq} \,
\lambda'_{bq} \, E_{\,{}_\Box}(x_f,x_T) \ + \ {\lambda'_{bq}}^2
\, E_{\,{}_\Box}(x_T,x_T)
\label{ckcb}
\end{eqnarray}
where we have introduced the definitions
\begin{equation}
\lambda_{U_i,D_j D_k} = {(V_{CKM}^\ast)}_{i j} {(V_{CKM})}_{i k} \ ,
\qquad \lambda'_{D_i D_j} = V^{(T)\ast}_i \, V^{(T)}_j \ , \qquad x_f =
\frac{m_f^2}{m_W^2}\ ,
\end{equation}
together with the previous associations $(D_1\ D_2\ D_3) = (d\ s\ b)$,
$(U_1\ U_2\ U_3) = (u\ c\ t)$. The Inami-Lim function $E_{\,{}_\Box}(x,y)$
is given by~\cite{Inami:1980fz}
\begin{equation}
E_{\,{}_\Box}(x,y) \ = \ \frac{4 \, -\, 7 x y}{4\,(x-1)\,(y-1)}
\; + \left[ \; \frac{x^2\, (4-8x+xy)\,\log x}{4\,(x-1)^2\,(x-y)}
\; + \; (\;x\;\to\; y\;) \; \right] \ .
\end{equation}
Since in our case the $V_{CKM}$ matrix is not unitary, we cannot introduce
the usual unitarity relations to take into account the GIM cancellations
in Eqs.~(\ref{ckcb}). However, owing to the unitarity of the $V_L^u$ and
$V_L^d$ rotation matrices, the following relation is found to hold:
\begin{equation}
\sum_{i=1,3} \lambda_{U_i,D_j D_k} \; + \; \lambda'_{D_j D_k} \ = \
\sum_{i=1,3} {V_0^{d\ast}}_{ij} \, {V_0^d}_{ik} \ =
\ - \, V_{Xj}^{d\ast} V_{Xk}^d \ .
\end{equation}
Using this relation the coefficients $C_K$, $C_{B_q}$ can be written as
\begin{eqnarray}
C_K & \simeq & \lambda_{c,sd}^2\,S(x_c,x_c)\, + \, \lambda_{t,sd}^2 \,
S(x_t,x_t)\, + \, 2 \, \lambda_{c,sd} \, \lambda_{t,sd} \, S(x_c,x_t)
\, \nonumber \\
& & + \, 2\, \lambda_{c,sd} \, \lambda'_{sd} \, S(x_c,x_T) \, + \, 2\,
\lambda_{t,sd} \, \lambda'_{sd} \, S(x_t,x_T) \, +
\, {\lambda'_{sd}}^2 \, S(x_T,x_T)\, \nonumber \\
& & - \, 2\, V_{Xj}^{d\ast} V_{Xk}^d \left[ \lambda_{c,sd}
(E_{\,{}_\Box}(0,x_c) - 1) \, + \, \lambda_{t,sd} (E_{\,{}_\Box}(0,x_t) -
1) \, + \, \lambda'_{sd} (E_{\,{}_\Box}(0,x_T) - 1)\right]
\label{ck} \\
C_{B_q} & \simeq & \lambda_{t,bq}^2 \, S(x_t,x_t) \nonumber \\
& & + \, 2\, \lambda_{t,bq} \, \lambda'_{bq} \, S(x_t,x_T) \, +
\, {\lambda'_{bq}}^2 \, S(x_T,x_T)\, \nonumber \\
& & - \, 2\, V_{Xj}^{d\ast} V_{Xk}^d \left[\, \lambda_{t,sd}
(E_{\,{}_\Box}(0,x_t) -
1) \, + \, \lambda'_{sd} (E_{\,{}_\Box}(0,x_T) - 1)\right]
\label{cb}
\end{eqnarray}
where $S(x,y)$ is the Inami-Lim function usually considered in Standard
Model calculations,
\begin{eqnarray}
S(x,y) & = & - \, \frac{3\, x\, y}{4\,(x-1)\,(y-1)}
\; + \left[ \; \frac{x\, y\, (4-8x+x^2)\,\log x}{4\,(x-1)^2\,(x-y)}
\; + \; (\;x\;\to\; y\;) \; \right] \ .
\end{eqnarray}
In the limit $x\to y$ one has~\cite{Buchalla:1995vs}
\begin{equation}
S(x,y) \ \to \ S_0(x) \ = \ \frac{4x - 11 x^2 + x^3}{4\,(x-1)^2} \; + \;
\frac{3\,x^3\,\log x}{2\,(x-1)^3} \ \ .
\end{equation}
In Eqs.~(\ref{ck}) and (\ref{cb}) we have taken the limit $x_u\to 0$, and
we have neglected the contributions to $C_{B_q}$ driven by $\lambda_c$ and
a term proportional to $(V_{Xj}^{d\ast} V_{Xk}^d)^2$. Notice that in both
equations the first line corresponds to the usual SM contribution, the
second line includes the contribution from exotic quarks and the third
line is a residual contribution that arises from the nonunitarity of the
$V_0^d$ matrix. Concerning this last term, it is worth to point out that
the experimental values of $\Delta m_K$ and $\Delta m_{B_q}$ also provide
constraints on the nondiagonal elements of $V_0^d$. Indeed, as shown in
Ref.~\cite{Cabarcas:2007my}, the latter lead to tree level FCNC mediated
by the $Z'$ gauge boson. In addition, in Eqs.~(\ref{ck}) and (\ref{cb}) we
have neglected perturbative QCD corrections. These are typically below a
30\% level~\cite{Buchalla:1995vs}, therefore they are not relevant in
order to estimate the order of magnitude of the bounds for the couplings
involving the $T$ quark.

Now, taking into account the experimental measurements for the $\Delta
m_P$ observables, and assuming that there is no fortunate cancellations
with other contributions from nonstandard physics (e.g.\ the mentioned
$Z'$-mediated FCNC), we can establish bounds for the products
$|\lambda'_{D_i D_j}| = |V^{(T)\ast}_i \, V^{(T)}_j|$, $ij = 12$, 13 and
23, for a given value of the $T$ quark mass. To do this, we cannot take
into account the usual estimations of SM contributions, which assume the
unitarity of $V_{CKM}$. Indeed, the top quark mixing angles in the SM are
basically determined from the same observables we want to use to constrain
the new parameters. Thus, in order to estimate upper bounds for the $T$
mixing angles, we will just take into account the $c$ quark contribution
to $\Delta m_K$ ($\lambda_{c,sd}$ can be measured from direct
observation), while the remaining SM contributions will be set to zero.

\begin{figure}[htb]
\vspace*{0.5cm}
\centerline{
   \includegraphics[height=6.9truecm]{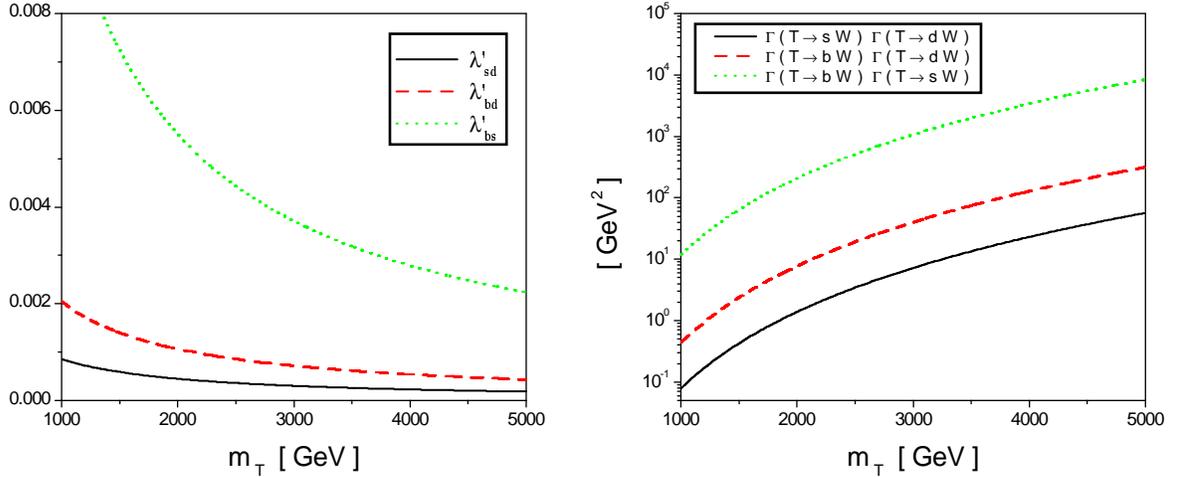}
   } \caption{Upper bounds for $|\lambda'_{D_i D_j}|$ (left panel) and
$\Gamma(\,T\ \rightarrow \ D_i\, W^+\,) \Gamma(\,T\ \rightarrow \ D_j\,
W^+\,)$ (right panel) as functions of the $T$ quark mass.}
\label{fig1}
\end{figure}

For the numerical analysis we will use the experimental
results~\cite{Yao:2006px,Abulencia:2006ze}
\begin{eqnarray}
\Delta m_K & = & m_{K_L} - m_{K_S} \ = \ (5.292\pm 0.009)\times 10^{-3}\
{\rm ps}^{-1} \nonumber \\
\Delta m_{B_d} & = & m_{B_H^0} - m_{B_L^0} \ = \ 0.507\pm 0.005\ {\rm
ps}^{-1} \nonumber \\
\Delta m_{B_s} &  = & m_{B_{sH}^0} - m_{B_{sL}^0} \ = \ 17.77\pm 0.12\
{\rm ps}^{-1} \ \ . \label{deltas}
\end{eqnarray}
Taking the central values of these measurements (errors are negligible at
the level of accuracy of our estimations) we obtain the results shown in
the left panel of Fig.~\ref{fig1}, where we plot the bounds for
$|\lambda'_{D_i D_j}|$ as functions of the $T$ quark mass. It can be seen
that the values obtained are of the order of $10^{-3}$, and they decrease
for increasing $m_T$. This can be understood by noting that $S_0(x_T)\sim
x_T/4$ for large values of $x_T$.
%%%%%%%%%%%%%%%%%%%%%%%%%%%%%%%%%%%%%%%%%%%%%%%%%%%%%%%%%%%%%%%%%%%%%%%
%%%%%%%%%%%%%%%%%%%%%%%%%%%%%%%%%%%%%%%%%%%%%%%%%%%%%%%%%%%%%%%%%%%%%%%
In addition, one can relate the bounds for $|\lambda'_{D_i D_j}|$ with the
corresponding bounds for the decay widths of the exotic $T$ quark into a
$W^+$ boson and a down-like quark $d$, $s$ or $b$. These widths are given
by
\begin{equation}
\Gamma(\,T\ \rightarrow \ D_i\, W^+\,) \ = \ \frac{G_F}{\sqrt{2}}\,
\frac{m_T^3}{8\pi} \, \left|V^{(T)}_i \right|^2 \,
\left(1\, - \, 3\, y_W^4 \, + \, 2 \, y_W^6\right)\ ,
\end{equation}
where $y_W \equiv \ m_W/m_T$. Since the $\Delta m_P$ observables involve
products $|V^{(T)\ast}_i \, V^{(T)}_j|$, one can establish upper bounds
for the products $\Gamma(\,T\ \rightarrow \ D_i\, W^+\,) \Gamma(\,T\
\rightarrow \ D_j\, W^+\,)$, for $i\neq j$. The corresponding numerical
results are shown in the right panel of Fig.~\ref{fig1}, where we plot
these bounds as functions of the exotic $T$ quark mass.

\begin{figure}[htb]
\vspace*{0.5cm}
\centerline{
   \includegraphics[height=6.9truecm]{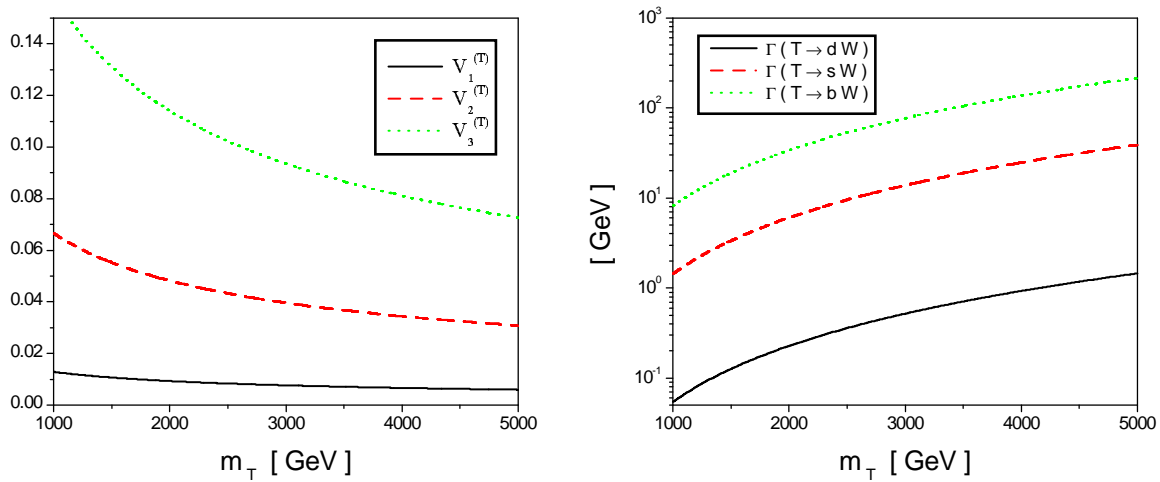}
   } \caption{Left: values of $|V^{(T)}_i|$ that simultaneously saturate
the experimental bounds of the observables $\Delta m_K$, $\Delta m_{B_d}$
and $\Delta m_{B_s}$. Right: $T \to D_i W^+$ decay widths that correspond
to the values in the left panel.}
\label{fig2}
\end{figure}

The experimental values of $\Delta m_P$ do not allow us to establish
separate bounds for the $|V^{(T)}_i|$ parameters. However, it is
interesting to consider the case in which all three experimental
constraints are saturated simultaneously. In this situation one finds the
values for $|V^{(T)}_i|$ shown in the left panel of Fig.~\ref{fig2} (as
before, we show the plots as functions of $m_T$). As expected, the
couplings between the exotic $T$ quark and the ordinary $d$, $s$ and $b$
quarks are suppressed according to the usual family hierarchy. Notice that
in principle one could have $Tb$ mixing angles as large as $\sim 0.1$ for
$T$ quark masses of a few TeV. Finally, from the values of $|V^{(T)}_i|$
one can immediately obtain the corresponding $T\to D_i W^+$ decay widths.
These are quoted in the right panel of Fig.~\ref{fig2}, as functions of
the $T$ quark mass. We notice that the dependence on $m_T$ vanishes for
the ratios between the decays into down-like quarks of different families,
the corresponding branching ratios obeying approximate relations
\begin{equation}
\frac{{\rm BR}(T\to d\,W^+)}{{\rm BR}(T\to s\,W^+)} \ \simeq \
\frac{1}{30} \ \ ,
\qquad \frac{{\rm BR}(T\to d\,W^+)}{{\rm BR}(T\to
b\,W^+)} \ \simeq \ \frac{1}{150}\ \ ,
\label{relbr}
\end{equation}
which arise from the ratios between $V^{(T)}$ matrix elements.
%%%%%%%%%%%%%%%%%%%%%%%%%%%%%%%%%%%%%%%%%%%%%%%%%%%%%%%%%%%%%%%%%%%%%%%
%%%%%%%%%%%%%%%%%%%%%%%%%%%%%%%%%%%%%%%%%%%%%%%%%%%%%%%%%%%%%%%%%%%%%%%

\hfill

Let us now analyze the $T$ decays into a $Z$ boson and an ordinary up-like
quark. From the currents in Eq.~(\ref{nct}) we have
\begin{eqnarray}
\Gamma(\,T\ \rightarrow \ t\, Z\,) & = & \frac{G_F}{\sqrt{2}}\,
\frac{m_T^3}{16\pi} \, |\lambda^Z_{Tt}|^2 \, \left(1 +
y_Z^2 - 2\,y_t^2 - 2\,y_Z^4 + y_Z^2\, y_t^2 + y_t^4\right)\,
\sqrt{[1-(y_Z + y_t)^2] [1-(y_Z-y_t)^2]} \nonumber \\
\Gamma(\,T\ \rightarrow \ U_i\, Z\,) & = & \frac{G_F}{\sqrt{2}}\,
\frac{m_T^3}{16\pi} \, |\lambda^Z_{TU_i}|^2 \,
\left(1\, - \, 3\, y_Z^4 \, + \, 2 \, y_Z^6\right) \ \ ,
\ \ i = 1,\ 2 \ \ ,
\end{eqnarray}
where $y_t = m_t/m_T$, $y_Z = m_Z/m_T$, and we have defined
\begin{equation}
\lambda^Z_{TU_i} \ \equiv \ \sum_{i=1}^3 V_{Xj}^{u\ast} V_{0\,ji}^u \ .
\end{equation}
In principle, both the order of magnitude of the matrix elements
$V^u_{Xj}$ and $V^u_{0\, ji}$ cannot be constrained independently from
present experimental measurements. However, in order to have an estimation
of the possible size of the decay widths, we can take into account the
values of $|V^{(T)}_i|$ obtained above, together with some assumptions on
the mixing matrices $V_0^{u,d}$. In this sense, most popular models of
quark mass matrices assume that off-diagonal elements of mixing matrices
satisfy family hierarchies given by
\begin{equation}
V^q_{0\,ij} \ \sim \ \left(\frac{m_{q_i}}{m_{q_j}}\right)^{1/2} \ ,
\label{assum}
\end{equation}
for $q = U$, $D$. From Eq.~(\ref{vt}), and taking into account
the results shown in the left panel of Fig.~2, one has then
\begin{eqnarray}
& & |\lambda^Z_{Tu}| \ \sim \ |V^u_{X1}| \ \sim \ 0.01 \nonumber \\
& & |\lambda^Z_{Tc}| \ \sim \ |V^u_{X2}| \ \sim \ 0.05 \nonumber \\
& & |\lambda^Z_{Tt}| \ \sim \ |V^u_{X3}| \ \sim \ 0.1 \ .
\end{eqnarray}
For the decay widths of the exotic $T$ quark into $U_i\, Z$ states,
$U_i=u$, $c$, $t$, one obtains approximately the same dependence on $m_T$
as in the case of $\Gamma(T\ \rightarrow \ D_i\, W^+)$ processes, together
with a global kinematical factor $\simeq 1/2$. Thus we have for each family
\begin{equation}
\Gamma(T\ \rightarrow \ D_i\, W^+) \  \ \simeq \  \,
2\, \Gamma(T\ \rightarrow \ U_i\, Z).
\label{widths}
\end{equation}

The above relations provide a couple of hints on what we can expect from
exotic $T$ decays if they are observed in future colliders: on one hand,
for any value of $m_T$ the branching ratios for $T\to U_i\, Z$ and $T\to
D_i W^+$ should be of the same order of magnitude, being $U_i$ and $D_i$
up and down quarks of the same family. On the other hand, the decay widths
should obey family hierarchies, as stated in Eq.~(\ref{relbr}). As stated,
these relations correspond to the situation in which the 331 contributions
saturate the bounds on the $\Delta m_P$ mass differences.

Let us recall that we have considered here the 331 model with $\beta = -
1/\sqrt{3}$, in which one has only one exotic quark $T$ with electric
charge $Q=2/3$. One can also study the model with $\beta=+1/\sqrt{3}$, in
which one has two exotic quarks of this kind, $T_{1,2}$. Though the
theoretical treatment remains qualitatively similar, in this case one has
to deal with more unknown parameters (masses and mixing angles), and the
phenomenological analysis is obscured. Therefore we have concentrated here
on the first possibility. Another important difference between both models
is that in the case $\beta = -1/\sqrt{3}$ the theory includes extra heavy
neutrinos, while for $\beta = +1/\sqrt{3}$ one has exotic charged leptons.

\section{$T\to qH$ decays}

Let us analyze the decays of a $T$ quark into an ordinary quark and a
scalar field. As stated in Sect.~II, the scalar eigenstates can be
separated into those with masses at the $\nu_\chi$ scale and a set of
fields that can be associated with the scalars of a THDM. Here we
concentrate on the decays of the $T$ into these lighter states, assuming
that the other channels are largely suppressed or even forbidden owing to
the large scalar masses.

The quark-scalar vertices arise from the Yukawa couplings in
Eq.~(\ref{yukawa-q}), which include a large number of unknown parameters.
Consequently, in order to get an estimation of the expected order of
magnitude of the relevant couplings it is necessary to rely on a definite
Ansatz for the quark mass matrices. A natural option in this sense is to
consider a scenario with approximate flavor symmetry such as that proposed
by Cheng and Sher~\cite{Cheng:1987rs,331case}, now extended to include the
heavy $T$ quark. This scenario is consistent with the assumption
introduced in the previous section, see Eq.~(\ref{assum}). Thus, for the
up-like quark sector we will write the coupling constants in
Eq.~(\ref{yukawa-q}) as
\begin{equation}
h^{i\Phi}_{U_j} \ = \ \lambda^{i\Phi}_{U_j} \sqrt{m_{U_i}
m_{U_j}}/\nu_\Phi\ ,
\end{equation}
with $\lambda^{i\Phi}_{U_j}\simeq {\cal O} (1)$. Within this Ansatz, the
dominant $T$ decays in the scalar sector arise from the terms driven by
the couplings $h_T^{m\rho}$ and $h_T^{3\eta}$. Now, if flavor symmetry is
approximately conserved, current quark eigenstates and mass quark
eigenstates are approximately the same. Let us identify the top and bottom
quarks with the quarks in the ${\bf 3}$ representation, i.e.\ $U_3$ and
$D_3$. Then the relevant couplings for the $T$ quark decays are
\begin{equation}
h_{T}^{3\eta } \left[ \, \bar t_LT_R\, (\cos\theta_\beta\, h^0 +
\sin\theta_\beta\, H^0 + i \sin\theta_\beta\, A^0 ) \
 + \ \bar b_L T_R \sin\theta_\beta\, H^- \,\right] \ .
\label{ht}
\end{equation}
Here we have neglected higher orders in $t-T$ mixing, which in the
framework of approximate flavor symmetry means to work at the leading
order in $(m_t/m_T)^2$. In this limit the corresponding decay widths are
given by
\begin{eqnarray}
\Gamma(\,T\ \rightarrow \ t\, h^0\,) & = & \frac{|h_T^{3\eta}|^2}{32\pi}
\,m_T\, \cos^2\theta_\beta \,\left(1 - y_{h^0} \right)^2
\nonumber \\
\Gamma(\,T\ \rightarrow \ q\, \phi\,) & = & \frac{|h_T^{3\eta}|^2}{32\pi}
\,m_T\, \sin^2\theta_\beta \,\left(1 - y_{\phi} \right)^2 \ ,
\label{widthshiggs}
\end{eqnarray}
where $y_X = (m_X/m_T)^2$, and in the second equation $q\,\phi\, = \,
tH^0,\ tA^0,\ bH^+$. If in addition we assume $\nu_\eta \simeq \nu_\rho$,
the global coupling constant $h_T^{3\eta}$ can be approximated by
$|h_T^{3\eta}|^2\;\simeq\; |\lambda^{3\eta}_T|^2\, 8\, m_t\, m_T\,
G_F/\sqrt{2}$.

As stated, we have identified the top and bottom quarks with those in the
${\bf 3}$ representation. This election is in principle arbitrary. If,
instead, we had chosen the $t$ and $b$ quarks to belong to one of the
families in the ${\bf 3^\ast}$, the results in Eqs.~(\ref{ht}) and
(\ref{widthshiggs}) would be qualitatively similar, with the interchange
$\cos\theta_\beta \leftrightarrow \sin\theta_\beta$,
$\eta\leftrightarrow\rho$. Since $\theta_\beta$ is an unknown parameter,
it is seen that the family choice is not relevant in order to obtain a
rough numerical estimation of the size of the decays.

To present some numerical results for the expected relative decay widths
of the $T$ quark, we will neglect the mass of the light Higgs boson $h^0$
compared with $m_T$, and we will take $m_H^0\simeq m_A^0\simeq m_H^\pm$
(in fact these masses are expected to be of the same order of magnitude,
see Table II). Finally, the sizes of $T\to tZ$ and $T\to bW^+$ decays will
be approximated taken into account the assumption in Eq.~(\ref{assum}),
which leads to $|\lambda^Z_{Tt}|^2\approx |V_3^{(T)}|^2\approx m_t/m_T$.
One gets in this way
\begin{equation}
\Gamma(T\to tZ)\ : \ \Gamma(T\to bW^+) \ : \ \Gamma(T\to th^0) \ :
\Gamma(T\to q\phi) \ \ \approx \ \
\frac{1}{2} \ : \ 1 \ : \ 2\cos^2\theta_\beta \ : \
2\sin^2\theta_\beta\left(1 - m_\phi^2/m_T^2\right)^2 \ \ ,
\label{relations}
\end{equation}
where, as before, $q\,\phi\, = \, tH^0,\ tA^0,\ bH^+$. Notice that in this
Cheng-Sher-like scenario the relative sizes of the decay widths do not
depend (at the leading order) on the $T$ quark mass. Only the phase space
factor $(1-m_\phi^2/m_T^2)$ appears in the case of $T\to q\phi$ decays.

Results from the relations in Eq.~(\ref{relations}) are shown in Table
III, where we quote approximate values for the branching ratios taking
different values for $\tan\theta_\beta$ and the intermediate scalar masses
$m_\phi$. We recall that one expects $m_\phi^2\sim |f|\nu_\chi$, where $f$
is a parameter that drives a trilinear coupling in the scalar potential.
The values in the table have been obtained after many assumptions and
approximations, therefore they should be taken only as illustrative.
However, since this corresponds to a plausible scenario, it is interesting
to notice that the decays into scalars might be the most important ones in
the search for an exotic $T$ quark.

\begin{center}
\begin{table}
\begin{tabular}
{c || c|c|c || c|c|c || c|c|c}
& \multicolumn{3}{c||}{$\tan\theta_\beta=0.1$} &
\multicolumn{3}{c||}{$\tan\theta_\beta=1$} &
\multicolumn{3}{c}{$\tan\theta_\beta=10$} \\
& $m_\phi\ll m_T$ & $m_\phi=\frac{m_T}{2}$ & $m_\phi\agt m_T$ & $m_\phi\ll
m_T$ & $m_\phi=\frac{m_T}{2}$ & $m_\phi\agt m_T$ & $m_\phi\ll m_T$ &
$m_\phi=\frac{m_T}{2}$ & $m_\phi\agt m_T$ \\
\hline
BR$(T\rightarrow b W^+)$
&0.14  & 0.14 & 0.15  &0.10   &0.12  & 0.20 & 0.06 & 0.10 &0.33\\
\hline
BR$(T\rightarrow t Z)$   &0.28 & 0.28 & 0.28  &0.18   &0.23  & 0.40 & 0.13
& 0.21 &0.66\\
\hline
BR$(T\rightarrow t h^0)$ &0.57  & 0.57
& 0.57  &0.18   &0.23  & 0.40 & 0.002 & 0.004 &0.01\\
\hline
BR$(T\rightarrow q \phi)$ ($\times$ 3) &0.015  & 0.01 & 0  &0.54
&0.42 & 0 & 0.81 & 0.69 &0\\
\hline
\end{tabular}
\caption{Approximate branching ratios for $T$ decays into scalars and
gauge bosons in a Cheng-Sher-like scenario.} \label{BRhh}
\end{table}
\end{center}

\section{Summary}

In summary, we have studied here the phenomenology of exotic $T$ quarks in
the framework of 331-symmetric models. We have concentrated on the models
with $\beta=-1/\sqrt{3}$, in which one has a single $T$ with charge 2/3
that in general mixes with the ordinary $u$, $c$ and $t$ quarks. We have
studied in detail the couplings of this $T$ quark with the ordinary gauge
bosons, establishing bounds for $T\to qW$ decays from the measured values
of neutral $K$, $B_d$ and $B_s$ mass differences. Then we have analyzed
the decays $T\to qZ$, considering the situation in which the previous
bounds are saturated, together with some assumptions on the hierarchies in
the quark mixing angles. As expected, the bounds are in agreement with
family hierarchies. The dependence with the $T$ quark mass is shown in
Figs.~1 and 2. Finally we have analyzed the decays of the $T$ quark into a
scalar and an ordinary quark. Though the results are strongly dependent on
model parameters that are in principle unknown, it is possible to present
some estimations for the widths by considering a definite scenario in
which one has approximate flavor symmetries. By performing plausible
assumptions on the order of magnitude of coupling constants and mass
scales, it can be seen that the decays into fermion-scalar states may be
indeed the dominant ones.

\section{Acknowledgments}

DGD thanks L.~Anchordoqui for useful discussions. This work has been
supported in part by CONICET and ANPCyT (Argentina, grants PIP 6009 and
PICT04-03-25374), COLCIENCIAS (Colombia), Fundaci\'on Banco de la
Rep\'ublica (Colombia), and the High Energy Physics
Latin-American-European Network (HELEN).

\end{document}